\documentclass[showpacs,preprintnumbers,aps,prd,a4paper,superscriptaddress,nofootinbib,eqsecnum]{revtex4} 
\usepackage{graphicx} 
\usepackage{latexsym} 
\usepackage{amsmath,amssymb}        

\renewcommand\({\left(}  
\renewcommand\){\right)}  
\renewcommand\[{\left[}  
\renewcommand\]{\right]}

\newcommand\ee{\end{equation}}  
\newcommand\be{\begin{equation}}  
\newcommand\eea{\end{eqnarray}}  
\newcommand\bea{\begin{eqnarray}}

\def\cala{{\cal A}}

\def\calr{{\cal R}}

\newcommand\bfk{{\bf k}}

\newcommand\g{\,\mbox{g}}  
  
\newcommand\GeV{\,\mbox{GeV}}

\newcommand\Mpc{\,\mbox{Mpc}}

\begin{document}  
  
\title{Constraints on the primordial curvature perturbation 
from primordial black holes}  
\author{Ignacio Zaballa} 
\affiliation{Cosmology and Astroparticle Physics Group, 
Department of Physics, University of Lancaster, Lancaster LA1 4YB, 
United Kingdom} 
\author{Anne M.~Green} 
\affiliation{School of Physics and Astronomy, University of Nottingham, 
Nottingham, NG7 2RD, United Kingdom} 
\author{Karim A.~Malik} 
\affiliation{Cosmology and Astroparticle Physics Group, 
Department of Physics, University of Lancaster, Lancaster LA1 4YB, 
United Kingdom} 
\author{Misao Sasaki} 
\affiliation{Yukawa Institute for Theoretical Physics, Kyoto University, 
Kyoto 606-8502, Japan} 
\date{\today} 
\pacs{98.80.Cq  \hfill  YITP-06-66, \,\,\, astro-ph/0612379} 
 
\begin{abstract} 
We calculate the constraints on the primordial curvature perturbation 
at the end of inflation from the present day abundance of Primordial 
Black Holes (PBHs), as a function of the reheat temperature $T_{\rm 
RH}$.  We first extend recent work on the formation of PBHs on scales 
which remain within the horizon during inflation and calculate the 
resulting constraints on the curvature perturbation. We then evaluate 
the constraint from PBHs that form, more conventionally, from 
super-horizon perturbations. The constraints apply for $T_{\rm RH} < 
10^{8} {\rm GeV}$ and the inclusion of sub-horizon PBHs leads to a 
limit which is roughly three times tighter than the bound  
from super-horizon PBHs. 
\end{abstract} 
 
\maketitle  
 
\section{Introduction} 
 
The spectrum of the primordial curvature perturbation is accurately 
measured on large scales by observations of the Cosmic Microwave 
Background (CMB) and Large Scale Structure (LSS) surveys. However, on 
small scales\footnote{Here and in the following by ``small scales'' we 
mean scales that are smaller than the horizon size at decoupling.} 
the only constraints come from avoiding the overproduction of 
Primordial Black Holes (PBHs).

Typically PBHs are light black holes which may form in the early 
Universe, in particular via the collapse of large density 
perturbations~\cite{Carr:1974nx,Carr:1975qj}. If the density 
perturbation at horizon entry in a given region exceeds a threshold 
value (which is of order unity) then the region will collapse to form 
a PBH with mass roughly equal to the horizon mass.  There are tight 
limits on the initial abundance of PBHs from the consequences of their 
expected evaporation via Hawking radiation and their present day 
abundance~\cite{const1,const2,const3,const4,const5,const6,const7,const8,const9,const10,const11} (for a review of these constraints see 
Ref.~\cite{carrrev}).  These limits can in turn be used to constrain 
the amplitude of the primordial perturbations on small scales and 
hence constrain models of inflation over a far wider range of scales 
than those constrained by CMB and LSS 
observations~\cite{Carr:1993,Carr:1994ar,Kim:1996hr,Green:1997sz,Leach:2000ea,CE}. 
To date these calculations have been carried out in terms of the 
density perturbation. In this paper we calculate, for the first time, 
the constraints in terms of the amplitude of the primordial curvature 
perturbation power spectrum. 
 
The standard PBH formation calculation applies to scales which are 
well outside the horizon at the end of inflation, with PBH formation 
subsequently occurring on a given scale, if the fluctuations are 
sufficiently large, when that scale re-enters the horizon after 
inflation.  Lyth et al.~\cite{Lyth:2005ze} extended the calculation to 
smaller scales on which the perturbations do not exit the horizon 
during inflation, and therefore never become classical. We use the 
sub-horizon PBH formation calculation of Ref.~\cite{Lyth:2005ze} to 
calculate the constraint on the primordial curvature perturbation from 
the present day abundance of sub-horizon PBHs and compare this with 
the conventional constraint from scales which exit the horizon during 
inflation.

Throughout this work we use `super-horizon' (`sub-horizon') to refer 
to scales which do (do not) exit the horizon during inflation. Note, 
however, that in the super-horizon case PBH formation occurs when the 
scale re-enters the horizon after inflation.  In Sec.~\ref{rev} we 
outline the calculation of the evolution of the Bardeen potential 
after the end of inflation during radiation domination from the 
initial conditions set during the inflationary epoch, and the formation 
of PBHs on sub-horizon scales. In Sec.~\ref{constraint_calc} we 
calculate the constraints on the primordial curvature perturbation 
from the present day density of both super- and sub-horizon PBHs 
before finally discussing our results in Sec.~\ref{discussion}.

\section{PBH formation on sub-horizon scales} 
\label{rev} 
 
During inflation the vacuum fluctuations of any light scalar field are 
promoted to a classical perturbation after horizon exit 
\cite{Polarski:1995jg,lyth06}.  The quantum to classical transition of 
the fluctuations can be described using the Heisenberg picture. In the 
Heisenberg picture the mode function of the field perturbations, 
$\delta\phi_k$, becomes real up to a constant phase transformation 
well after horizon crossing (see Eq.(\ref{mode_funct}) below), and 
therefore the time evolution of the corresponding operator 
$\delta\hat\phi_{\bf k}(t)=\delta\phi_k(t) \hat a_{\bf k} + 
\delta\phi_k^{\star}(t) \hat a^{\dagger}_{-\bfk}$, where 
$\hat a_\bfk$ and  
$\hat a_\bfk^{\dagger}$ are the annihilation and creation operators 
respectively, is trivial ensuring that it has a steady eigenstate over 
subsequent times. 
For fluctuations on scales that never exit the horizon during inflation,  
the evolution of the field operators, $\delta\hat\phi_\bfk(t)$, can not be 
regarded as classical because in this case the field operators do not commute  
at different times. 
However, if the perturbations are sufficiently large, then PBH formation  
could still occur at the end of inflation on sub-horizon scales if we  
regard the formation of a black hole itself as a measurement of the  
overdense region. 
In Ref.~\cite{Lyth:2005ze} it was shown that the formation of sub-horizon 
PBHs in this context should be formulated in terms of the Bardeen potential  
$\Phi$, rather than using, for example, the density contrast.  To find the  
abundance of sub-horizon PBHs we therefore need to calculate the Bardeen 
potential immediately after inflation, for scales which remain sub-horizon  
during inflation.  This was done in Ref.~\cite{Lyth:2005ze}. We summarise  
here the essential details of the calculation. 
 
\subsection{Evolution of the Bardeen potential} 
\label{Bardeen_rev} 
 
We write the metric with scalar perturbations 
in the longitudinal (or Newtonian) gauge as~\cite{Kodama:1985bj} 
\be 
ds^2=a^2\left(\tau\right)\left[ 
-\left(1+2\Psi\right)d\tau^2+\left(1+2\Phi\right)\delta_{ij}dx^idx^j 
\right]  \,. 
\ee   
In this gauge the metric perturbations coincide with the gauge 
invariant Bardeen potentials defined in~\cite{Bardeen:1980kt}, 
$\Psi=\Phi_A$ and $\Phi=\Phi_H$.  
For a scalar field the anisotropic stress is zero so that the Bardeen 
potentials are related by $\Phi = -\Psi$ and $\Psi$ is related to 
$\delta$, the comoving density contrast by~\cite{Bardeen:1980kt,Kodama:1985bj} 
\be  
\delta_{k} =- \frac{2}{3} \left(\frac{k}{aH} \right)^2 \Psi_{k}\,,  
\ee  
where $a$ is the scale factor, $k$ the comoving wave number, $H$ the 
Hubble parameter defined as $H=\dot a/a$ where the dot denotes 
differentiation with respect to coordinate time $t$, and we denote 
Fourier components by a subscript ``$k$''. 
If inflation is driven by a light slowly rolling field $\phi$ then, to 
leading order in the slow roll approximation, the field equation for 
field fluctuations, $\delta \phi$, on flat slices lives in unperturbed 
spacetime. Defining $\psi \equiv a \delta \phi$, the Fourier components 
of the field perturbations obey  
\be 
\frac{{\rm d}^2 \psi({\bf k}, \tau)}{{\rm d} \tau^2} 
+ \left( k^2 - \frac{2}{\tau^2} \right) \psi({\bf k, \tau}) = 0 \,, 
\ee 
where $\tau=-(a H)^{-1}$ is conformal time (it is assumed that inflation 
is exponential). The solution of this equation, with initial condition 
corresponding to the flat space-time mode function $\psi(k, \tau) 
= \exp{(-i k \tau)}/\sqrt{2 k}$, 
is 
\be\label{mode_funct} 
\psi(k,\tau) = \frac{1}{\sqrt{2 k}} (1 + i k \tau) \frac{1}{ik \tau}  
         \exp{(-i k \tau)} \,. 
\ee 
The Fourier components of the comoving curvature perturbation ${\cal 
R}$ are related to the field fluctuations on flat slices $\delta \phi$ 
by 
\be  
{\cal R}_{k} = - \frac{H}{\dot{\phi}} \delta \phi_{k} \,. 
\ee 
For a definition of ${\cal R}$ in terms of the metric perturbations 
and its relation to curvature perturbations defined on different 
hypersurfaces see for example Ref.~\cite{kam_thesis}.  
On sub-horizon scales therefore 
\be 
\label{eq:Rinfl} 
{\cal{R}}_{\rm{k}}\left(t\right)=  
-\frac{1}{\sqrt{2k^3}}\frac{H^2}{\dot\phi}\left(i+\frac{k}{aH}\right){\rm{exp}} 
\left(\frac{ik}{aH}\right)\,. 
\ee 
The Bardeen potential $\Psi$ is related to $\calr$ by 
\begin{equation} 
\label{eq:Rconservation} 
\frac{2}{3H}\dot{\Psi}_k+\frac{\left(5+3w\right)}{3}\Psi_k 
=-\left(1+w\right){\cal{R}}_k\,, 
\end{equation} 
where $w=p/\rho$, with $p$ and $\rho$ being the pressure and the 
energy density, respectively. Under slow-roll conditions during 
inflation $w\simeq -1$, and therefore $\Psi$ is practically zero.

We need to set the initial conditions for the radiation dominated 
epoch after the end of inflation. We assume for simplicity that the 
slow-roll conditions hold at the end inflation and that reheating is 
rapid. This is not necessarily the case; the decay and thermalization 
of the inflaton may be slow and in single field models slow-roll has 
to be violated at the end of inflation.  Our assumptions are valid in, 
for instance, hybrid inflation models where the secondary waterfall 
field has a large mass so that the slow-roll conditions hold at the 
end of inflation, and the inflaton field rolls to the minimum of its 
effective potential at the end of the inflationary stage almost 
instantaneously~\cite{Linde:1993cn}.

Under these conditions, we can smoothly match the solutions of the 
Bardeen potential between inflation and radiation domination by 
requiring the intrinsic metric and the extrinsic curvature to be 
continuous on comoving 
hypersurfaces~\cite{Israel:1966rt,Deruelle:1995kd,Martin:1997zd}. 
Then $\Psi$ during the radiation dominated epoch 
is given by~\cite{Lyth:2005ze} 
\be 
\label{psisol} 
\Psi_{k}(\tau) = \frac{3 ( 1 + w) {\cal R}_{k}(\tau_{\rm e})}{2 x^3}  
    \left[ (x - x_{\rm e}) \cos{(x-x_{\rm e})} -  
    (1 + x x_{\rm e}) \sin{(x-x_{\rm e})} \right] \,, 
\ee 
where  
\be 
\label{def_x} 
x \equiv c_{\rm s} k \tau \,, 
\ee 
the subscript ``$e$'' denotes the end of inflation, and $x_{\rm e} 
\equiv c_{\rm s} k \tau_{\rm e} = c_{\rm s} k/ a_{\rm e} H_{\rm 
e}$. On super-horizon scales, $k \ll a_{\rm e} H_{\rm e}$, and 
Eq.(\ref{psisol}) reduces to the standard relation.  On sub-horizon 
scales the Bardeen potential undergoes damped oscillations, reaching a 
maximum value $\Psi(x_{\star})$ at $x=x_{\star}$ during the first 
oscillation.  The mass variance of $\Psi$ is approximately given by 
$\sigma_{\Psi}^2(M) \approx {\cal P}_{\Psi}(M)$ so that 
\be 
\sigma^2_{\Psi}(M,t) = \frac{4 {\cal P}_{\cal R}(x_{\rm e})}{{x}^6} 
      \left[ ({x}-x_{\rm e}) \cos{({x}-x_{\rm e})} 
    - (1+ {x} x_{\rm e}) \sin{({x}-x_{\rm e})} \right]^2 \,, 
\ee 
where, for $x_{\rm e} \gtrsim c_{\rm s}$, 
\be 
{\cal P}_{\cal R}(x_{\rm e})  
= {\cal A}_{\cal R}   
\left[ 1 + \frac{x_{\rm e}^2}{c_{\rm s}^2} \right] \,, 
\ee 
defining ${\cal A}_{\cal R} \equiv [(H_{\rm e}^2/\dot{\phi}_{\rm e})/2 
\pi]^2$, the amplitude of the curvature perturbation power spectrum at 
the end of inflation on super-horizon scales.

In Ref.~\cite{Lyth:2005ze}, it was argued that Eq.(\ref{psisol}) is no 
longer valid when the amplitude of the curvature perturbation $\calr$ 
becomes greater than one.  Since ${\cal R}$ increases with $(k/a H)$, 
there would then be a scale below which it appears that linear theory 
breaks down.  At the end of inflation this scale is given by 
$(k/a_{\rm e}H_{\rm e})\sim1/\calr_{\rm HC}$, where $\calr_{\rm HC}$ 
is the amplitude of $\calr$ at horizon crossing.  However this is a 
conservative limit because well inside the horizon gravity should be 
less important, and therefore it is the scalar field perturbation 
becoming larger than unity which signals the break down of linear 
theory.  
An estimate of the critical wavenumber can be found by equating 
the energy density of the field perturbation and 
the background energy density. This gives   
$(k/a_{\rm e})_{\rm crit}\sim\sqrt{M_{\rm P}H_{\rm e}}$,  
which can be written in terms of the reheating temperature as 
\be\label{philimit} 
\(\frac{k}{a_{\rm e}H_{\rm e}}\)_{\rm crit} 
\sim5\times 10^{10}\times\(\frac{10^7\GeV}{T_{\rm RH}}\)\,. 
\ee  
The critical wavenumber decreases with increasing reheating 
temperature.  The minimum value, which occurs for a reheating 
temperature of $10^{16}$GeV, is the same order of magnitude as the 
critical wavenumber corresponding to $\calr \sim 1$ if $\calr_{\rm HC} 
\sim 0.01-0.1$. For smaller reheat temperatures this criterion gives a 
larger critical wavenumber than $\calr \sim 1$. We discuss the 
consequences for the constraint on the amplitude of the curvature 
perturbation of these two limits in Sec.~\ref{constraint_calc}.

\subsection{PBH abundance} 
\label{formation_rev} 
 
%
In the standard super-horizon calculation of the abundance of PBHs one 
applies the Press-Schechter formalism, which is often used in 
large-scale structure studies \cite{PS}. The quantity of interest 
(usually the density field, but in this application the Bardeen 
potential) is smoothed on a mass scale $M$, and the regions where the 
field perturbations exceed a certain threshold value are assumed to 
form PBHs with mass greater than M. The fraction of the Universe in 
PBHs with a mass similar to or larger than the smoothing 
scale\footnote{We assume throughout that the PBH mass is equal to the 
smoothing scale mass, which in the standard, super-horizon case 
corresponds to the horizon mass $M_{{\rm H}}$. This is not strictly 
true, and numerical simulations indicate that the PBH mass depends on 
the size and shape of the perturbations~\cite{nj,ss,musco}; however, 
this uncertainty is not important when applying PBH abundance 
constraints, due to their relatively weak mass dependence.}, is given 
by Press-Schechter theory, initially, as 
\be 
\label{omegaps} 
\Omega_{{\rm PBH, i}}(>M) 
\equiv  
F(M) =  
2 \int_{|\Psi_{\rm c}|}^{\infty} P(\Psi(M)) \,   {\rm d}  \Psi(M)  
=   {\rm erfc}  
\left( \frac{|\Psi_{\rm c}|}{\sqrt{2}\sigma^{\star}_\Psi(M)} \right) \,,  
\ee 
where $\sigma^{\star}_\Psi(M)$ is the maximum mass variance of $\Psi$, 
and the final equality follows from the fact that the smoothed Bardeen 
potential $\Psi(M)$ has a Gaussian probability distribution.  We have 
followed the usual Press-Schechter practise of including a factor of 
``2'' in Eq.~(\ref{omegaps}). The collapse criterion in terms of this 
variable is $\vert\Psi\vert> \vert\Psi_{\rm c}\vert = 1/2$, which is 
equivalent to a density contrast threshold at horizon crossing 
$\delta_{\rm c} =1/3$.   
The mass fraction of PBHs per logarithmic mass 
interval is 
\be 
M \frac{{\rm d} F(M)}{{\rm d} M}  
    =  - \sqrt{ \frac{2}{\pi}}  
   \frac{M|\Psi_{\rm c}|}{{\sigma^{\star2}_{\Psi}}} 
       \frac{{\rm d} \sigma^{\star}_{\Psi}}{{\rm d} M}  
  \exp{ \left( -\frac{\Psi_{\rm c}^2}{2 \sigma^{\star2}_{\Psi}} \right)}\,. 
\ee 
In the short wave-length limit, 
$x_{\rm e} \gg 1$, $\Psi_k$ behaves like a damped sinusoidal function, 
and therefore the maximum occurs at $x_{\star}=x_{\rm e} + 
(\pi/2)$. In general there is no simple analytic 
solution for $x_{\star}$.   
The difference $B(x_{\rm e})=(x_{\star}-x_{\rm e})$ converges 
very rapidly to the value $\pi/2$ inside the sound  horizon.  
We used this fact in Ref.~\cite{Lyth:2005ze} 
to give an estimate of the mass fraction of PBHs. Here in order to extend 
the calculation to scales of the order of the horizon scale  
at the end of inflation, we calculate $B(x_{\rm e})$ numerically. 
This calculation breaks down when the first maximum occurs outside 
the horizon, $x_{\star} < c_{\rm s}$. 
The PBH mass is 
related to $x_{\rm e}$ by 
\be 
\label{PBHmass} 
M= \frac{4 \pi}{3} \rho_{\star} \left( \frac{a_{\star}}{k} \right)^3 
         = \frac{4 \pi}{3} \rho_{\star} \left( \frac{a_{\rm e}}{k} \right)^3 
     \frac{x_{\rm e}}{x_{\rm e} + B(x_{\rm e})} = M_{\rm e} 
    \frac{c_{\rm s}^3}{x_{\rm e}^2 (x_{\rm e} + B(x_{\rm e}))} \,, 
\ee 
where $M_{\rm e}$ is the horizon mass at the end of inflation, 
\be 
\label{me} 
M_{\rm e} = \frac{4 \pi}{3} \rho_{\rm e} ( H_{\rm e}^{-1})^3  
  = M_{\rm eq} \left(\frac{ g_{\rm eff}^{\rm eq}}{g_{\rm eff}^{\rm e}}  
   \right)^{1/3} \left( \frac{a_{\rm e}}{a_{\rm eq}} \right)^2 = 
   1.2 \times 10^{17} 
     {\rm g} \left( \frac{10^{7} \, {\rm GeV}}{T_{\rm RH}} \right)^2 \,, 
\ee 
with $T_{\rm RH}$ the reheat temperature at the end of inflation and 
$g_{\rm eff}$ the number of relativistic degrees of freedom; 
$g_{\rm eff}^{\rm eq} \approx 3$ and at high temperatures $g_{\rm eff} 
\sim 100$.

\section{Constraints} 
\label{constraint_calc} 

PBHs with $M_{\rm PBH} \gtrsim 5 \times 10^{14} {\rm g}$ will not have 
evaporated by the present day (their lifetime is longer than the age 
of the Universe) and their density can not be too large. Traditionally 
`not too large' was interpreted as $\Omega_{{\rm PBH}, 0} \equiv 
\rho_{\rm PBH, 0} / \rho_{\rm tot, 0}< 1$, where the subscript ``$0$'' 
denotes the present epoch. This constraint can be improved to 
$\Omega_{{\rm PBH}, 0} < \Omega_{{\rm cdm}, 0} \approx 0.47$, where 
the numerical value is the 3-$\sigma$ upper limit on the CDM density 
extracted from WMAP~\cite{spergel}, however this factor of two 
improvement has an essentially negligible effect on the constraints on 
the primordial perturbations due to the exponential dependence of the 
PBH abundance on the size of the perturbations.

Since $\rho_{\rm PBH} \propto a^{-3}$ and $\rho_{\rm tot} \propto 
a^{-3}$ ($\rho_{\rm tot} \propto a^{-4}$) during matter (radiation) 
domination, $\Omega_{{\rm PBH}, 0}= \Omega_{\rm PBH, eq}$ and 
\begin{equation} 
\label{rhopbh1} 
\rho_{\rm PBH, eq} = \int_{5 \times 10^{14} {\rm g}}^{\infty} 
                 M \frac{{\rm d} n_{\rm eq}}{{\rm d} M} {\rm d} M = 
           \int_{5 \times 10^{14} {\rm g}}^{\infty} 
                 M  \frac{{\rm d} n_{\rm i}}{{\rm d} M} 
  \left( \frac{a_{\rm i}(M)}{a_{\rm eq}} \right)^3 {\rm d} M \,, 
\end{equation} 
where the initial PBH number density ${\rm d}n_{\rm i}/{\rm d}M$ 
is given by 
\be 
\label{dndm} 
\frac{{\rm d} n_{\rm i}}{{\rm d} M} = \frac{\rho_{i}(M)}{M}  
     \frac{{\rm d} F(M)}{{\rm d} M} \,. 
\ee 

The initial total (radiation) density is given by 
\begin{equation} 
\label{rhorad} 
\rho_{\rm i} = \frac{\pi^2}{30} g_{\rm eff}^{\rm i} T_{\rm i}^4 
           = \rho_{\rm rad, eq} \left( \frac{g_{\rm eff}^{\rm i} } 
           {g_{\rm eff}^{\rm eq}} \right)  
      \left( \frac{T_{\rm i}}{T_{\rm eq}} \right)^4 
    = \frac{\rho_{\rm tot, eq}}{2} \left( \frac{g_{\rm eff}^{\rm i} } 
           {g_{\rm eff}^{\rm eq}} \right)  
      \left( \frac{T_{\rm i}}{T_{\rm eq}} \right)^4 \,. 
\end{equation} 
Combining Eqs.~(\ref{dndm}),(\ref{rhopbh1}) and (\ref{rhorad}), and using 
$a \propto g_{\rm eff}^{-1/3} T^{-1}$, 
the present day mass fraction of sub-horizon PBHs becomes  
 \begin{equation} 
\label{presentday} 
\Omega_{{\rm PBH}, 0}= \frac{1}{2} 
  \left( \frac{g_{\rm eff}^{\rm eq}}{ g_{\rm eff}^{\rm i}} \right)^{1/3} 
   \int_{5 \times 10^{14} {\rm g}}^{\infty}  
 \left( \frac{a_{\rm eq}}{a_{\rm i}} \right) \frac{{\rm d}F}{{\rm d} M} 
  {\rm d} M \,. 
\end{equation} 
The upper mass limit of the integral in Eq.~(\ref{presentday}) should 
correspond to a mass approximately given by the horizon mass at the 
end of inflation. However, the mass function decreases so rapidly for 
large masses that in practise the upper limit can be safely taken to 
infinity.  Using Eqs.~(\ref{PBHmass}) and (\ref{me}) the present day 
PBH mass fraction, can be expressed in terms of $x_{\rm e}$ as 
\be 
\label{omega_pbh1}  
\Omega_{\rm PBH,0}=7.2\times10^{8}\(\frac{T_{\rm RH}}{\GeV}\)  
\int^{x_{\rm 0}}_{0}\frac{x_{\rm e}}{x_{\rm e}+B(x_{\rm e})} 
     \frac{dF}{dx_{\rm e}}dx_{\rm e} \,,  
\ee  
where $x_0$ is the value of $x_{\rm e}$ corresponding to 
$M=5\times10^{14}\g$ which can be found from Eq.~(\ref{PBHmass}). For 
small reheating temperatures, $T_{\rm RH} \ll 10^{5} {\rm GeV}$,  
we have $B\left(x_0\right) \ll x_{0}$, and $x_0$ is given by 
\be  
x_0^3\simeq46.2\times\(\frac{10^7\GeV}{T_{\rm RH}}\)^2 \,.  
\ee  
For larger reheat temperatures $B(x_{0})$ can not be neglected, and 
therefore we calculate $x_0$ and hence $B(x_{0})$ iteratively using 
Eq.~(\ref{PBHmass}). 
 
\begin{figure} 
\includegraphics[angle=0,width=0.5\textwidth]{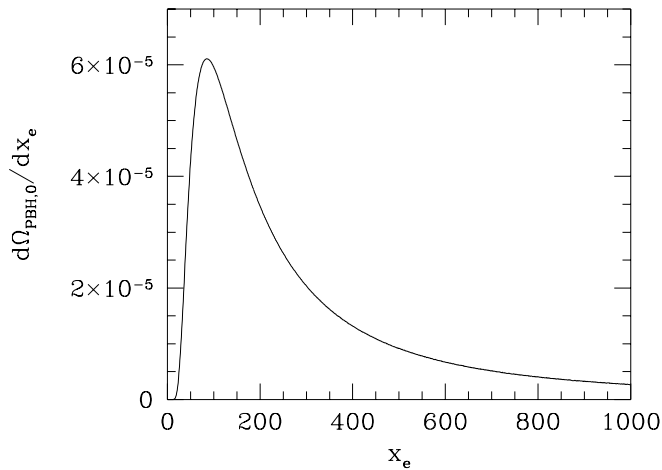} 
\caption{\label{fig:1} 
The differential mass fraction of sub-horizon PBHs, eq.~(\ref{omega_pbh1}),  
as a function of $x_{\rm e}$ 
for $\cala^{1/2}=0.02$ and $T_{\rm RH}= 1$ GeV.  
The distribution is sharply peaked at  
small $x_{\rm e}$, demonstrating that 
the majority of PBHs form at essentially the same time.} 
\end{figure}

\begin{figure}  
\includegraphics[angle=0,width=0.5\textwidth]{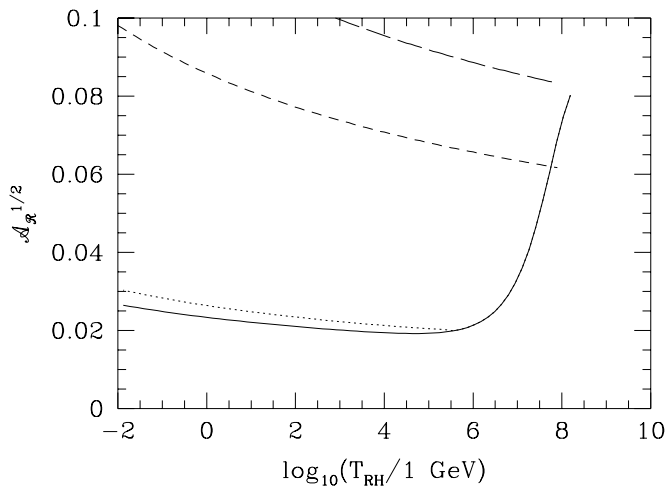}  
\caption{\label{fig:2}  
The constraint on the amplitude of the curvature perturbation 
power spectrum 
at the end of inflation from the present day abundance of PBHs as a  
function of reheat temperature $T_{\rm RH}$. The solid and dotted lines are 
for sub-horizon PBHs. The solid line assumes that the linear theory 
calculation is still valid on scales on which the comoving curvature perturbation 
becomes larger than one before the end of inflation, while the dotted line 
ignores PBH formation on these scales (and hence provides a conservative 
evaluation of the constraint). 
The long and short dashed lines are for super-horizon PBHs using a density 
contrast threshold  $\delta_{\rm c}=1/3$ and $0.45$ respectively.}  
\end{figure}

In Fig.~\ref{fig:1} we plot the present day differential mass fraction 
of sub-horizon PBHs, ${\rm d} \Omega_{\rm PBH,0}/{\rm d} x_{\rm e}$ from 
eq.~(\ref{omega_pbh1}), as  
a function of $x_{\rm e}$ for ${\cal A}_{\cal R}^{1/2}=0.02$ 
and $T_{\rm RH}= 1$ GeV.   
The PBH differential mass fraction 
 is relatively sharply peaked with the majority of PBHs  
forming at essentially the same time. The present day PBH mass fraction, 
Eq.~(\ref{omega_pbh1}) is therefore given, to a good approximation, by 
\be 
\label{omega_pbh2}  
\Omega_{\rm PBH,0}\simeq7.2\times10^{8}\(\frac{T_{\rm RH}}{\GeV}\) 
     \int^{x_0}_0\frac{dF}{dx_{\rm e}}dx_{\rm e}  
    \simeq7.2\times10^{8}\(\frac{T_{\rm RH}}{\GeV}\)\[{\rm erfc} 
   \(\frac{|\Psi_{\rm c}|}{\sqrt{2}\sigma_\Psi^*(x_0)}\)\]\,,  
\ee  
for $x_{0} \gg 1$, and by  
\be 
\label{omega_pbh3}  
\Omega_{\rm PBH,0}  
\simeq7.2\times10^{8}\(\frac{T_{\rm RH}}{\GeV}\)\[\(\frac{x_0}{x_{\star}}\) 
   {\rm erfc}\(\frac{|\Psi_{\rm c}|}{\sqrt{2}\sigma_\Psi^*(x_0)}\)\]\,,  
\ee  
for $x_{0} \lesssim 1$ where, in both cases, $\sigma_\Psi^*(x_0)$ is the maximum 
value of the mass variance at $x_{\rm e}=x_0$. 
For a range of reheating temperature values, $T_{\rm RH}$, we 
calculate the constraint on ${\cal A}_{\cal R}$ by first calculating 
$x_{0}$, the value of $x_{\rm e}$ corresponding to PBHs with 
$M=5\times10^{14}\g$ (the lightest PBHs which will not have evaporated 
by the present day). We then find the mass variance of the 
Bardeen potential for this value of $x_{0}$, $\sigma_\Psi^*(x_0)$, 
using Eq.~(\ref{psisol}). Finally, we use either 
Eq.~(\ref{omega_pbh2}) or (\ref{omega_pbh3}), as appropriate, to 
calculate the constraint on ${\cal A}_{\cal R}$ from the requirement 
$\Omega_{\rm PBH, 0} < 0.47$. The results are plotted in 
Fig.~\ref{fig:2}. As the reheat-temperature increases the constraint 
initially becomes tighter as the duration of the radiation dominated 
era, during which the fraction of the energy density in the form of 
PBHs grows as $\rho_{\rm PBH} / \rho_{\rm tot} \propto a$, 
increases. As the reheat temperature increases the horizon mass at the 
end of inflation decreases and for $10^{6} \, {\rm GeV} < T_{\rm RH} < 
10^{8} \, {\rm GeV}$ only a fraction of the sub-horizon PBHs have $M > 
5 \times 10^{14} {\rm g}$.  For $T_{\rm RH} > 10^{8} \, {\rm GeV}$, 
$M_{\rm e} \sim 5 \times 10^{14} {\rm g}$ and none of the sub-horizon 
PBHs are massive enough to last to the present day.  
Up to this point 
we have ignored the fact, as pointed out at the end of 
Sec.~\ref{Bardeen_rev}, that the calculation assumes that the 
linear theory holds.  
If we ignore PBHs formed on scales where ${\cal R} > 1$ at the end of  
inflation, we see in Fig.~\ref{fig:2} that this weakens the 
constraints on ${\cal A}_{\cal R}^{1/2}$ for $T_{\rm RH} < 10^5 \, 
{\rm GeV}$, but the change is small (at most $20\%$). 
This is because the PBH mass function decreases rapidly for large $x_{\rm e}$. 
Taking the linear limit for the field 
perturbations in Eq.~(\ref{philimit}) the change in the constraint  
is negligible for the reheating temperatures for which our present day 
abundance constraint is applicable ($T_{\rm RH}< 10^{8}$ GeV).

We now compare the sub-horizon constraints with the conventional 
super-horizon constraints. A full calculation would require the 
complete curvature perturbation power spectrum, or equivalently the 
specification of a concrete inflation model. In order to be as general 
as possible we simply consider the amplitude of the super-horizon 
curvature perturbation power spectrum, ${\cal A}_{\cal R}$, as a free 
parameter. On cosmological scales ${\cal A}_{\cal R}=2.4 \times 
10^{-9}$~\cite{spergel}, therefore the production of an interesting 
(i.e.~non-negligible) density of PBHs requires  
the curvature perturbations to be significantly 
larger on small scales. Assuming that the power spectrum grows 
monotonically with increasing wavenumber the abundance of 
super-horizon PBHs will be dominated by PBHs which form immediately 
after inflation at $T_{\rm RH}$. The present day abundance of 
super-horizon PBHs is then given by 
\begin{equation} 
\Omega_{\rm PBH, 0} (M> 5 \times 10^{14} {\rm g}) 
             = 7.2. \times 10^{8} \frac{T_{\rm RH}}{{\rm GeV}} 
             {\rm erfc} \left( \frac{\delta_{\rm c}}{\sqrt{2}  
     \sigma_{\delta}(T_{\rm RH})} \right) \,. 
\end{equation} 
The Fourier components of the density contrast and the curvature 
perturbation are related by \cite{LL} 
\be 
\label{delr} 
\delta_{k}(t) = \frac{2(1+w)}{5+3w} \left( \frac{k}{aH} \right)^2  
        {\cal R}_{k} \,, 
\ee 
so that, at horizon crossing, ${\cal P}_{\delta} = (4/9)^2 {\cal 
P}_{\cal R} = (4/9)^2 {\cal A}_{\cal R}$. Using $\sigma_{\delta}(M) 
\approx {\cal P}_{\delta} (M)$ again, then $\sigma_{\delta}(M)= 
(4/9)^2 {\cal A}_{\cal R}$.  The threshold density has historically 
been taken as $\delta_{\rm c} = w$, where $w=p/\rho$ is the equation 
of state \cite{Carr:1975qj}, so that for formation during radiation 
domination (which is usually the case of interest) $\delta_{\rm 
c}=1/3$. More recently Green et al.~\cite{Green:2004wb} carried out a 
new calculation using peaks theory and a formation criterion derived 
by Shibata and Sasaki~\cite{ss} which refers to the central value of 
the metric perturbation. They found that the standard Press-Schechter 
calculation agrees with the peaks theory calculation if the density 
threshold contrast is in the range $0.3$ to $0.5$, while numerical 
simulations by Musco et al. found a threshold $\delta_{\rm 
c}=0.45$~\cite{musco}. We therefore consider two density thresholds, 
$\delta_{\rm c}=1/3$ and $0.45$.

The constraints on the curvature perturbation at the end of inflation, 
as a function of reheat temperature, are plotted in 
Fig.~\ref{fig:2}. The super-horizon constraints are a factor of $~\sim 
3-4$ weaker than the sub-horizon constraints, and the factor of $\sim 
1.4$ difference in the density contrast thresholds we consider gives, 
as would naively be expected, a similar difference in the constraints 
(with the larger threshold leading to a weaker constraint). The 
super-horizon constraint stops at $T_{\rm RH} \sim 10^{8} \, {\rm 
GeV}$ as then $M_{\rm e}< 5 \times 10^{14} {\rm g}$ and the PBHs which 
form immediately after inflation have evaporated by the present day. 
PBHs with $M> 5 \times 10^{14} {\rm g}$ could form later, at $T< 
T_{\rm RH}$, though their abundance would depend on the precise scale 
dependence of the power spectrum.

\section{Discussion} 
\label{discussion} 
 
CMB and LSS observations probe primordial perturbations on scales of 
the order $1\Mpc$ to $3000\Mpc$ that typically leave the horizon 
roughly 50-55 e-foldings before the end of inflation.  The amplitude 
and scale dependence of the perturbations on these scales are now 
determined fairly accurately through observations, 
however these scales correspond to only a 
small region of the inflationary potential. On the other hand
PBHs constrain the 
primordial fluctuations over a large range of far smaller scales, and 
hence provide additional and complementary
constraints on the dynamics of inflation over a wider 
range of energy scales. These constraints are relevant for models where 
the perturbations have significant scale-dependence, for instance 
models where there are multiple stages of inflation or 
inflation ends via a phase transition.

Observational constraints on the abundance of PBHs are usually carried  
out for scales which exit the horizon during inflation (with PBH  
formation occurring after horizon re-entry during the subsequent 
radiation dominated era).  Lyth et al.~\cite{Lyth:2005ze} recently 
studied the formation of PBHs on smaller scales on which the 
perturbations do not exit the horizon during inflation, and therefore 
never become classical. We have extended this work by calculating the 
constraints on the amplitude of the primordial curvature perturbation 
power spectrum at the end of inflation, ${\cal A}_{\cal R}$, from the 
present day abundance of these sub-horizon PBHs, as a function of the 
reheat temperature after inflation.  For $T_{\rm RH} < 10^{6} \, {\rm 
GeV}$ most of the sub-horizon PBHs have $M> 5 \times 10^{14} \, {\rm 
g}$ and hence have lifetimes longer than the age of the Universe. The 
requirement that their present-day density does not exceed the upper 
limit on the cold dark matter density leads to the constraint ${\cal 
A}_{\cal R} \lesssim 0.25-0.3$ with a weak dependence on the 
reheat-temperature (the higher the reheat-temperature, the longer the 
duration of the radiation dominated epoch during which the fraction of 
the energy density in PBHs grows). As the reheat temperature increases 
the horizon mass at the end of inflation decreases. For $10^{6} \, 
{\rm GeV} < T_{\rm RH} < 10^{8} \, {\rm GeV}$ the fraction of the 
sub-horizon PBHs which have $M > 5 \times 10^{14} {\rm g}$ decreases 
rapidly with increasing $T_{\rm RH}$ and hence the constraint on 
${\cal A}_{\cal R}$ is weakened. For $T_{\rm RH} > 10^{8} \, {\rm 
GeV}$, $M_{\rm e} \sim 5 \times 10^{14} {\rm g}$ so that none of the 
sub-horizon PBHs are massive enough to last to the present day and 
hence there is no constraint on ${\cal A}_{\cal R}$ from the present 
day abundance of sub-horizon PBHs. 
 
We also calculated the constraint from the present day abundance of
PBHs which form on larger scales, that do exit the horizon during
inflation. This is a factor of $\sim 3-4$ weaker than the constraint
from sub-horizon scales.For higher reheat temperatures there will be
constraints from the effects of the PBH evaporation products on the
successful predictions of
nucleosynthesis~\cite{const2,const3,const4,const5,const6,const7,const8,const9,const10,const11carrrev}. These
constraints are however model dependent, depending on the PBH mass
function and hence the scale-dependence of the primordial power
spectrum, which is beyond the scope of this paper.
 
There may also be constraints from the present day density of relics 
which may be the end point of PBH evaporation and the emission of 
long-lived massive particles.  The applicability, or otherwise, of 
these constraints depends on the details of physics beyond the 
standard model of particle physics and are hence also beyond the remit 
of this paper.

\acknowledgments 
 
The authors would like to thank David Lyth for useful discussions and 
comments. AMG and KAM were supported by PPARC.  
MS was supported by JSPS Grant-in-Aid for 
Scientific Research (S) No.~14102004, (B) No.~17340075, and (A) No.~18204024. 
 
  

\end{document}